\pdfoutput=1
\documentclass[a4paper,11pt]{article}
\usepackage[utf8]{inputenc}
\usepackage{jcappub}
\usepackage[utf8]{inputenc}
\usepackage{amsmath,amssymb}
\usepackage{graphicx}
\usepackage{multicol}
\usepackage{hyperref}
\usepackage{physics}
\usepackage{lipsum}
\usepackage{dcolumn}
\usepackage{tabularx}

\usepackage{pgfplots}
\usepgfplotslibrary{fillbetween}
\usetikzlibrary{patterns}
\pgfplotsset{compat=1.8}

\newcommand{\be}{\begin{equation}} 
\newcommand{\ee}{\end{equation}}
\newcommand{\beq}{\begin{equation}} 
\newcommand{\eeq}{\end{equation}}
\newcommand{\bea}{\begin{equation}\begin{aligned}} 
\newcommand{\eea}{\end{aligned}\end{equation}}
\newcommand{\ba}{\begin{eqnarray}}
\newcommand{\ea}{\end{eqnarray}}
\newcommand{\ie}{{\it i.e.} }

\newcommand{\Ri}{\mathcal{R}}

\usepackage[babel,german=quotes]{csquotes}

\title{\centering Palatini-Higgs inflation with non-minimal derivative coupling}

\author[a]{Ioannis D.~Gialamas,}
\author[b]{Alexandros Karam,}
\author[c]{Angelos Lykkas,}
\author[d]{and Thomas D.~Pappas}

\emailAdd{i.gialamas@phys.uoa.gr}
\emailAdd{alexandros.karam@kbfi.ee} 
\emailAdd{a.lykkas@uoi.gr}
\emailAdd{thomas.pappas@physics.slu.cz} 
 
\affiliation[a]{National and Kapodistrian University of Athens, Department of Physics,
Nuclear and Particle Physics Section,   GR--157 84   Athens,~Greece }
\affiliation[b]{Laboratory of High Energy and Computational Physics, 
National Institute of Chemical Physics and Biophysics, R{\"a}vala pst.~10, Tallinn, 10143, Estonia}
\affiliation[c]{Physics Department, University of Ioannina, GR–45110 Ioannina, Greece}
\affiliation[d]{Research Centre for Theoretical Physics and Astrophysics, Institute of Physics, Silesian University in
Opava, Bezruovo nm.~13, CZ-746 01 Opava, Czech Republic}

\abstract{
The predictions of standard Higgs inflation in the framework of the metric formalism yield a tensor-to-scalar ratio $r \sim 10^{-3}$ which lies well within the expected accuracy of near-future experiments $ \sim 10^{-4}$.  When the Palatini formalism is employed, the  predicted values of $r$ get highly-suppressed $r\sim 10^{-12}$ and consequently a possible non-detection of primordial tensor fluctuations will rule out only the metric variant of the model. On the other hand, the extremely small values predicted for $r$ by the Palatini approach constitute contact with observations a hopeless task for the foreseeable future. In this work, we propose a way to remedy this issue by extending the action with the inclusion of a generalized non-minimal derivative coupling term between the inflaton and the Einstein tensor of the form $m^{-2}(\phi) G_{\mu\nu}\nabla^{\mu}\phi \nabla^{\nu}\phi$. We find that with such a modification, the Palatini predictions can become comparable with the ones obtained in the metric formalism, thus providing ample room for the model to be in contact with observations in the near future.
}

\begin{document}

\maketitle

\vspace{1.5cm}

\section{Introduction}

Higgs inflation~\cite{Bezrukov2008, Bezrukov2009a, Rubio:2018ogq} is one of the simplest and most natural scenarios that can successfully describe the quasi-exponential expansion of the Universe in its very early stages. Since the Higgs boson is the only fundamental scalar field that has been observed in Nature, one may wonder if it can also play the role of the inflaton. In order to comply with the observational constraints \cite{Akrami:2018odb}, the Higgs must have a non-minimal coupling with gravity\footnote{Note that, in general, even if it is absent at tree-level, a non-minimal coupling will be generated from quantum corrections~\cite{Callan:1970ze}.}. However, in non-minimally coupled theories an issue arises as to which variational principle should be used.

In the metric formulation of gravity, the spacetime manifold is endowed with a metric $g_{\mu\nu}$ and a connection $\Gamma^{\rho}_{\mu\nu}$ that is Levi-Civita, i.e.~is metric compatible and torsion-free. This allows the connection to be uniquely determined in terms of the metric and its derivatives and thus the latter is effectively the only independent degree of freedom (DOF).  On the other hand, in the Palatini formulation~\cite{Palatini1919, Ferraris1982} (also encountered in the literature as \enquote{metric-affine} or \enquote{first-order} formalism) the metric and the connection are treated as independent DOF. Their dynamics are governed by a set of field equations that stem from the independent variations of the action with respect to both fields and, in principle, ${\Gamma^{\rho}}_{\mu\nu}$ will not be of the Levi-Civita type. In the context of the standard theory of General Relativity (GR) the two formulations turn out to be equivalent since they yield the same field equations, with the main difference being that the Levi-Civita connection is recovered on-shell in the Palatini formalism. This ceases to be the case though in more elaborate theories when for example higher-curvature terms are taken into account and/or matter couples non-minimally with gravity~\cite{Exirifard2008, Bauer2008, Bauer2011, Tamanini2011, Bauer:2010jg, Olmo2011, Rasanen2017, Tenkanen:2017jih, Racioppi2017, Markkanen:2017tun, Jaerv2018, Fu:2017iqg, Racioppi2018, Carrilho:2018ffi, Kozak:2018vlp, Bombacigno2019, Enckell2019, Rasanen2019, Antoniadis2018, Rasanen2018, Almeida2019, Antoniadis2019, Shimada2019, Takahashi2019, Jinno2019, Tenkanen2019, Edery2019, Rubio:2019ypq, Jinno2020, Aoki2019, Giovannini2019, Tenkanen2019a, Bostan2019a, Bostan2019, Tenkanen2020, Gialamas2020, Racioppi2019, Antoniadis2019a, Tenkanen2020a, Tenkanen2020b, Shaposhnikov:2020fdv, LloydStubbs2020, Antoniadis2020, Borowiec:2020lfx, Ghilencea2020, Das:2020kff, Jarv:2020qqm, Gialamas:2020snr, Karam:2020rpa, McDonald:2020lpz, Langvik:2020nrs, Ghilencea:2020rxc, Shaposhnikov:2020gts, Shaposhnikov:2020frq}. Consequently this fact serves as motivation to study extensions of GR within the framework of the Palatini formalism.

A straightforward approach in extending GR consists of postulating the existence of additional DOF in the form of scalar fields that interact non-minimally with the gravity sector of the action. Such modifications belong in the general class of the so-called scalar-tensor (ST) theories (see for example \cite{Clifton:2011jh} and references therein) that play a prominent role in the study of early-Universe Cosmology since they provide a natural setup for the description of the inflationary phase with the scalar field being identified with the inflaton.

One may further extend the ST class of theories by considering non-minimal derivative couplings (NMDC) of the inflaton with the curvature \cite{Amendola:1993uh} (see also~\cite{Amendola:1993uh, Capozziello:1999uwa, Capozziello:1999xt, Germani:2010gm, Amendola:2010bk, Tsujikawa:2012mk, Sadjadi:2012zp, Kamada:2012se, Koutsoumbas:2013boa, Luo:2014eda, Germani:2014hqa, Hohmann:2015kra, Ema:2015oaa, Gumjudpai:2015vio, Zhu:2015lry, Myung:2016twf, Harko:2016xip, Dalianis:2016wpu, Goodarzi:2016iht, Tumurtushaa:2019bmc, Granda:2019wpe, Granda:2019jqy, Oliveros:2019xef, Dalianis:2019vit, Fu:2019ttf, Granda:2019wip, Sato:2020ghj, Bayarsaikhan:2020jww, Oikonomou:2020sij, Gao:2020jhq}). For example, in the case of terms that contain four derivatives the possible combinations are $\kappa_1 R\nabla_{\mu}\phi\nabla^{\mu}\phi$, $\kappa_2 R_{\mu\nu}\nabla^{\mu}\phi \nabla^{\nu}\phi$, $\kappa_3 R \phi\square\phi$, $\kappa_4 R_{\mu\nu} \phi\nabla^{\mu}\nabla^{\nu}\phi$, $\kappa_5 \nabla_{\mu}R \phi \nabla^{\mu}\phi$ and $\kappa_6 \square R \phi^2$ ,where the coupling constants $\kappa_1,\dots,\kappa_6$ have dimensions of $[\text{mass}]^{-2}$. Using total divergences and without loss of generality it has been shown that the terms $\kappa_1 R\nabla_{\mu}\phi \nabla^{\mu}\phi$ and $\kappa_2 R_{\mu\nu}\nabla^{\mu}\phi\nabla^{\nu}\phi$ alone suffice to encapsulate the properties of these theories. In general, the NMDC terms for arbitrary values of the coupling constants $\kappa_1$ and $\kappa_2$ yield third-order field equations. However, in the case of $\kappa_2=-2\, \kappa_1$ the field equations are of second-order \cite{Sushkov:2009hk} avoiding in this way the Ostrogradsky instability that is associated with the emergence of ghosts \cite{Ostrogradsky:1850fid,Woodard:2015zca}. Under this constraint for the coupling constants, the allowed fourth-derivative NMDC corrections of the action morph into a single term corresponding to a coupling between the Einstein tensor and the derivatives of the scalar field that upon a further promotion of the coupling to an arbitrary function of $\phi$ can be generalized to $m^{-2}(\phi)G_{\mu\nu}\nabla^{\mu}\phi \nabla^{\nu}\phi$ \cite{Dalianis:2019vit}. 

In order to study the inflationary predictions of ST theories in the absence of NMDC terms, the usual approach is to first recast the action in the Einstein frame (EF) where the Einstein-Hilbert term is decoupled from  the scalar field. The transformation of the general action to the EF is achieved by means of a Weyl rescaling of the metric $\widetilde{g}_{\mu\nu}= \Omega^2(\phi) g_{\mu\nu}$, also known as a \textit{conformal transformation}\footnote{A conformal transformation refers to a change in the coordinates, however we adopt the convention of the community where a Weyl rescaling and a conformal transformation are used interchangeably.}. In general, when NMDC terms are included in the action the coupling functionals of the Lagrangian will depend on both the scalar field and its canonical kinetic term $X \equiv- \frac{1}{2} \nabla^{\alpha}\phi \nabla_{\alpha}\phi$. Consequently, the Weyl rescaling in this case becomes inadequate and has to be replaced by the more general \textit{disformal transformation} of the metric $\widetilde{g} _{\mu\nu}= \Omega ^2(\phi,X) \left[ g_{\mu\nu} +\beta ^2(\phi,X) \nabla _\mu \phi\, \nabla _\nu \phi \right]$ that was originally proposed by J. Bekenstein in~\cite{Bekenstein:1992pj} (see also~\cite{Zumalacarregui:2013pma,Minamitsuji:2014waa,Tsujikawa:2014uza,Domenech:2015hka,Sakstein:2015jca,vandeBruck:2015tna,Achour:2016rkg,Gumjudpai:2016ioy,Escriva:2016cwl,Lobo:2017bfh,Sato:2017qau,Galtsov:2018xuc, Karwan:2018eln,Delhom:2019yeo,Galtsov:2020jnu,Qiu:2020csx}) and were shown to leave the Horndeski action invariant, just as conformal transformations leave the ST action invariant. 
Note that for $\beta ^2(\phi,X) = 0$ the disformal transformation reduces to the usual Weyl rescaling of the metric.

In this work, we investigate the predictions of Higgs inflation in the presence of the aforementioned generalized NMDC term within the framework of the Palatini formalism. With the use of a disformal transformation we bring the action to the Einstein frame where the inflationary observables can be readily computed. In this process, non-standard canonical terms are generated, which can modify the predictions of the model. While in the standard version of Palatini Higgs inflation the tensor-to-scalar ratio $r$ is predicted to be very small, i.e.~$\mathcal{O}(10^{-12})$~\cite{Bauer:2008zj, Rasanen2017, Markkanen:2017tun, Takahashi2019}, we find that the NMDC allows us to raise the value of $r$ considerably, even above $r \sim 10^{-4}$, a range that will be probed by future experiments such as LITEBIRD \cite{Matsumura2016}, PIXIE \cite{Kogut_2011}, and PICO \cite{Sutin:2018onu}. 

The paper is organized as follows. In Sec.~\ref{sec:palatini_Higgs} we provide an overview of the Higgs inflationary model in the Palatini formulation and express the inflationary observables in terms of the number of $e$-folds and the model parameters. Then, in Sec.~\ref{sec:NMDC} we consider an ST theory augmented with the addition of an NMDC term in the Palatini formalism. By employing a disformal transformation we bring the action to the Einstein frame. After that, in Sec.~\ref{sec:examples} we assume the potential to be of the Higgs type and focus on two cases for the NMDC: i) $m^2 = \text{const.}$ and ii) $m^2 \propto \phi^2$. We study the phenomenology of these cases and compare the predictions with the standard Palatini Higgs inflation. We conclude in Sec.~\ref{sec:conclusions} and present some analytic expressions for the inflationary observables in the Appendix.

\vspace{0.75cm}

\section{Overview of Palatini-Higgs inflation}
\label{sec:palatini_Higgs}

One of the conceptually simplest realizations of inflation is the Higgs boson to assume the role of the inflaton field. In the unitary gauge, the dynamics of the theory can be effectively described by the following action (see~\cite{Bauer2008})
\begin{equation} \label{eq:SJordan}
	S = \int \dd^4 x \sqrt{-g} \bigg[  \frac{1}{2}\qty(M^2 + \xi \phi^2 )g^{\mu\nu}R_{\mu\nu}\left[\Gamma,\partial\Gamma\right] - \frac{1}{2}g^{\mu\nu}\nabla_\mu\phi\nabla_\nu\phi - \frac{\lambda}{4}\phi^4 \bigg] \, .
\end{equation}
We can safely assume that at inflationary scales near the Planck scale, the Higgs field, denoted here by $\phi$, takes up values far away from its vacuum expectation value. Clearly, we can assume the scalar field $\phi$ to be an additional SM field and not necessarily the Higgs field. Here $M$ is a mass scale to be identified with the Planck scale $M_\text{Pl}$ (we assume $M_\text{Pl}\equiv1$ throughout this work) and $\xi$ is the non-minimal coupling of $\phi$ to gravity.

By eliminating the non-minimal coupling term, through a Weyl rescaling of the form
\begin{equation} \label{eq:metric_recaling}
	g_{\mu\nu}(x) \rightarrow \Omega^{-2}(\phi)g_{\mu\nu}(x) \, ,
\end{equation}
where
\begin{equation} \label{eq:Omega}
	\Omega(\phi) = \sqrt{1 + \xi \phi^2} \,,
\end{equation}
we obtain the action in the Einstein frame
\begin{equation} \label{eq:SEinstein1}
	S = \int \dd^4 x \sqrt{-g} \bigg[ \frac{1}{2}g^{\mu\nu}R_{\mu\nu}\left[\Gamma,\partial\Gamma\right] - \frac{1}{2\Omega^2(\phi)}g^{\mu\nu}\nabla_\mu\phi\nabla_\nu\phi - \frac{\lambda \phi^4}{4\Omega^4(\phi)} \bigg] \,.
\end{equation}
Notice that in the Palatini formulation the $R_{\mu\nu}$ is explicitly independent of the metric $g_{\mu\nu}$ and therefore remains unaffected by the Weyl rescaling.

It will prove useful to make a field redefinition $\phi\mapsto\chi$ in order to have a canonical kinetic term for the scalar field; that reads as
\begin{equation} \label{eq:inflaton_redef}
\frac{\mathrm{d}\phi}{\mathrm{d}\chi}=\sqrt{1+\xi\phi^2}\,,
\end{equation}
which leads to
\be 
\chi = \frac{1}{\sqrt{\xi}} \sinh^{-1} \left( \sqrt{\xi}\, \phi \right)
\quad \iff \quad
\phi = \frac{1}{\sqrt{\xi}} \sinh(\sqrt{\xi}\,\chi) \, .
\ee
Then, the Einstein frame action for the inflaton reads
\begin{equation} \label{eq:SE_infl}
	S = \int \dd^4 x \sqrt{-g} \qty[ \frac{1}{2}g^{\mu\nu}R_{\mu\nu}\left[\Gamma,\partial\Gamma\right] - \frac{1}{2}g^{\mu\nu}\nabla_\mu\chi\nabla_\nu\chi - U(\chi) ] \, ,
\end{equation}
with the self-interacting potential now given by
\be \label{eq:U}
U(\chi) = \frac{\lambda}{4\xi^2} \tanh^4\qty(\sqrt{\xi}\chi) \, .
\ee
At large field values the potential tends to a plateau allowing us to describe inflation in accordance to observational constraints.

Assuming a flat FLRW background [our metric convention is $\eta_{\mu\nu}=\text{diag}(-,+,+,+)$], the equations of motion are
\begin{equation} \label{eq:FRW}
	3H^2 = \frac{\dot{\chi}^2}{2} + U(\chi) \, , \qquad \qquad \ddot{\chi} + 3H\dot{\chi} + U'(\chi) = 0 \, ,
\end{equation}
where dot and prime denote derivatives with respect to cosmic time $t$ and the function's argument (here $\chi$), respectively. In the slow-roll approximation they become:
\begin{equation} \label{eq:SR}
	3H^2 \approx U \, , \qquad \qquad 3H\dot{\chi} + U' \approx 0 \, .
\end{equation}
The duration of inflation is measured by the number of $e$-folds
\begin{equation} 
    N = \int_{\chi_{\rm end}}^{\chi_*} \frac{U(\chi)}{U'(\chi)} \dd \chi \approx \frac{1}{16 \xi} \cosh \left( 2 \sqrt{\xi} \chi_* \right) \approx \frac{\phi_*^2}{8} \, .    
\end{equation}
The validity and most of the dynamics of the slow-roll approximation are encoded in the slow-roll parameters  
\begin{equation}
    \epsilon \simeq \frac{1}{8\xi N_*^2} \, , \qquad \qquad
    \eta \simeq -\frac{1}{N_*} \, .
\end{equation}
Both of them are small ($\ll 1$) during inflation and one of them approaches unity near its end. In terms of the slow-roll parameters, the observable quantities measured in the CMB are given by  
\begin{equation}
    A_s=\frac{1}{24\pi^2}\,\frac{U_*}{\epsilon_*},\qquad \quad n_s=1-6\epsilon_*+2\eta_*,\qquad\quad r=16\epsilon_*
\end{equation}
and are the power spectrum of scalar perturbations, the scalar spectral index and the tensor-to-scalar ratio, respectively. Their values are calculated at the horizon crossing, where $\phi=\phi_*$, as indicated by the star subscript in their expressions. Their observational bounds set by the Planck collaboration~\cite{Akrami:2018odb} are 
\begin{equation}\label{eq:Planck_constr}
    A_s\simeq2.1\times10^{-9},\qquad\quad n_s=\left\{\begin{matrix}\left(0.9607,0.9691\right),&\ 1\sigma\text{ region}\\\left(0.9565,0.9733\right),&\ 2\sigma\text{ region}\end{matrix}\right.,\qquad\quad r\lesssim0.056.
\end{equation}
Assuming an expansion around large $N_*$ we obtain the following expressions:
\begin{equation} \label{eq:cmb_observables}
    A_s \approx \frac{\lambda N_*^2}{12 \pi^2\xi} \,, \qquad \quad
n_s \approx 1 - \frac{2}{N_*} \,, \qquad \quad
r \approx \frac{2}{\xi N_*^2} \,,
\end{equation}
From the measured value of $A_s$ we obtain the relation
\begin{equation} \label{eq:pert_norm_couplings}
	\xi \approx 4 \times 10^6 N_*^2 \lambda \, ,
\end{equation}
implying that there is only one free parameter $\xi$ or $\lambda$. Without the running of the Higgs self-coupling $\lambda$, there is no way to determine the value of the parameters near the inflationary scale. Assuming a conservative value of $N_*=50-55$ $e$-folds the parameters lie in the following range:
\begin{equation}
    \xi\in\left[10^5,10^9\right]\quad \Longleftrightarrow\quad \lambda\in\left[10^{-5},10^{-1}\right]\,.
\end{equation}
Therefore in the Palatini formulation a large value of $\xi$ is needed, which in turn suppresses the tensor-to-scalar ratio $r\sim10^{-12}$, in contrast with the metric formalism where $r\sim10^{-3}$. Future missions~\cite{Matsumura2016, Kogut_2011, Sutin:2018onu} are expected to probe the region of $10^{-4}$ of $r$ and therefore models claiming predictions in that region can be hopefully distinguished.

\vspace{0.75cm}

\section{Non-minimal derivative coupling}
\label{sec:NMDC}

It has been shown \cite{Capozziello:1999xt, Sushkov:2009hk} that in order to consider derivative couplings between gravity and matter, it is enough to examine only the terms $\Ri^{\mu\nu} \nabla _\mu \phi \nabla _\nu \phi$ and $\Ri \nabla ^\mu \phi  \nabla _\mu \phi $ without loss of generality (see also \cite{Amendola:1993uh, Capozziello:1999uwa}). Respecting that statement we consider the following action:

\ba\label{action1_p}
S = \int \dd^4x \sqrt{-\widetilde{g}} \left( \frac{f(\phi)}{2} \widetilde{\Ri} -\frac{1}{2}\widetilde{\nabla}_\mu \phi \widetilde{\nabla}^\mu \phi
+\frac{1}{2 m^2(\phi)}\widetilde{G}^{\mu\nu}\widetilde{\nabla}_\mu \phi \widetilde{\nabla}_\nu \phi  -V(\phi) \right)\,, 
\ea
where $\widetilde{G}^{\mu\nu}[\widetilde{g},\Gamma]$ is the Einstein tensor given by $\widetilde{G}^{\mu\nu} = \widetilde{\Ri}^{\mu\nu} -\frac{1}{2}\widetilde{g}^{\mu\nu} \widetilde{\Ri}\,.$ The inflaton field $ \phi $ is non-minimally coupled with gravity via the Ricci scalar $ \widetilde{\Ri} $  and the Einstein tensor $\widetilde{G}^{\mu\nu}\,,$ by the functions $f(\phi)$ and $m^2(\phi)$ respectively.  In order to consider the Palatini formulation of gravity the Ricci scalar is defined as $ \widetilde{\Ri} = \widetilde{g}^{\mu\nu} \Ri_{\mu\nu}(\Gamma)\,, $ \ie the Riemann tensor is constructed only from the connection $ \Gamma\,, $ thus it is independent of the metric tensor. 

\subsection{Use of the disformal transformation}

In order to study the inflationary dynamics of the theory, we need to rephrase the action in the Einstein frame. In the standard Palatini-Higgs inflation of Sec.~\ref{sec:palatini_Higgs} this is easily implementable using the conformal transformation \eqref{eq:metric_recaling}. In our case a more general transformation is needed; the so-called disformal transformation which is given by
\be\label{disf_trans_p}
\widetilde{g} _{\mu\nu}= \Omega ^2(\phi,X) \left[ g_{\mu\nu} +\beta ^2(\phi,X) \nabla _\mu \phi\, \nabla _\nu \phi \right] \,.
\ee
where $X=-\frac{1}{2} \nabla _\mu \phi \nabla ^\mu \phi\,, $ is the canonical kinetic term of the field. After using the disformal transformation \eqref{disf_trans_p}, the action \eqref{action1_p} can be written in the manageable form
\begin{align}\label{action1_p2_m}
S_D =& \int\! \mathrm{d}^4x\,\sqrt{-g} \Bigg[ F_1(\phi,X)\frac{\Ri}{2} -F_2(\phi,X)\frac{ (\nabla \phi)^2}{2}  + F_3(\phi,X) \Ri_{\mu\nu} \nabla ^\mu \phi \nabla^\nu \phi   \nonumber
\\ &\qquad\qquad\qquad +F_4(\phi,X)\frac{ (\nabla \phi)^4}{4} - F_5(\phi,X)\, V(\phi) \Bigg]\,,
\end{align}
where the subscript D denotes the resulting action after the disformal transformation. The functions $F_i$ are given by
\ba
F_1(\phi,X) &=& f(\phi) \Omega ^2 \sqrt{1+\varepsilon u^2} -\frac{1}{2m^2(\phi)} \frac{\varepsilon u^2/\beta ^2}{\sqrt{1+\varepsilon u^2}}\,,
\\ F_2(\phi,X) &=&  \Omega ^2 \sqrt{1+\varepsilon u^2}\,,
\\ F_3(\phi,X) &=&  -\frac{f(\phi)}{2} \frac{\Omega ^2 \beta ^2}{\sqrt{1+\varepsilon u^2}} + \frac{1}{4m^2(\phi)} \frac{2 +\varepsilon u^2}{(1+\varepsilon u^2)^{3/2}}\,,
\\ F_4(\phi,X) &=&  \frac{2\Omega ^2 \beta ^2}{\sqrt{1+\varepsilon u^2}}\,, 
\\ F_5(\phi,X) &=&  \Omega ^4 \sqrt{1+\varepsilon u^2}\,,\label{F_func_m}
\ea
where $ \varepsilon u^2 = u_\mu u^\mu=\beta ^2 (\nabla \phi)^2 $. In order to obtain the action in the Einstein frame, we essentially have to solve the system
\be\label{system_m}
F_1(\phi,X) = 1   \qquad \text{and} \qquad  F_3(\phi,X) = 0 \,,
\ee
which results in obtaining the solutions for the transformation functions $\Omega ^2 $ and $\beta ^2$ as functions of the field and its velocity. The solution of \eqref{system_m} is easily obtained and reads
\ba\label{omega_beta_m} 
\Omega ^2&=& \frac{2+ \varepsilon u^2}{2 f(\phi) \sqrt{1+\varepsilon u^2}}\qquad \text{and} \qquad \beta ^2 =\frac{1}{m^2(\phi) \sqrt{1+\varepsilon u^2}}\,.
\ea
After substituting \eqref{omega_beta_m} to \eqref{action1_p2_m}-\eqref{F_func_m} we obtain the Einstein-frame action
\be
S_E = \int \mathrm{d}^4x \sqrt{-g} \left[ \frac{\Ri}{2} -\hat{F}_2(\phi,X) \frac{ (\nabla \phi)^2}{2}  + \hat{F}_4(\phi,X) \frac{ (\nabla \phi)^4}{4}  - \hat{F}_5(\phi,X)\,V(\phi)\right]\,,\label{action_Einstein_m}
\ee
with
\ba
\hat{F}_2(\phi,X) &=& \frac{2+ \varepsilon u^2}{2 f(\phi)}\,, \label{f2hat_m}
\\ \hat{F}_4(\phi,X) &=&  \frac{2+ \varepsilon u^2}{f(\phi) m^2(\phi) (1+\varepsilon u^2)^{3/2} }\,,\label{f4hat_m}
\\ \hat{F}_5(\phi,X) &=&  \frac{(2+ \varepsilon u^2)^2}{4 f^2(\phi) \sqrt{1+\varepsilon u^2}}\,. \label{f5hat_m}
\ea
Notice that although we have managed to recast the action into the Einstein frame, the prefactors of the kinetic terms and the potential are functionals that involve both the field $\phi$ and its canonical kinetic term $X$. To this end we need to take a further step and separate the $\phi$ and $X$ dependence of these terms before moving on to the computation of the inflationary observables. Using the canonical kinetic term $X$ and substituting $u^2=2\beta ^2 X$ in \eqref{omega_beta_m}, we find that
\be\label{u_X_m}
u^2(1-u^2)^{1/2}=\frac{2X}{m^2(\phi)}\,,
\ee
where we have used that $\varepsilon = -1$. As far as the inflationary dynamics are concerned, the kinetic term $X$ can be ignored, since the field is slowly rolling in that era. Therefore, an expansion around small values of $X$ is justified. Then, expanding eq. \eqref{u_X_m} in terms of $X\,,$ we find that
\ba\label{expansion:u_m}
u^2 \simeq \frac{2X}{m^2(\phi)}\left( 1+\frac{X}{m^2(\phi)}\right)\,.
\ea
Substituting \eqref{expansion:u_m} in \eqref{f2hat_m}-\eqref{f5hat_m} and keeping terms up to $\mathcal{O}\left(X^2\right)$ we obtain
\ba
\hat{F}_2(\phi,X) &\simeq& \frac{1}{ f(\phi)}\left(1-\frac{X}{m^2(\phi)} -\frac{X^2}{m^4(\phi) } \right)\,, \label{f2hat:ap_m}
\\ \hat{F}_4(\phi,X) &\simeq&  \frac{2}{ f(\phi) m^2(\phi)}\left(1+\frac{2X}{m^2(\phi)} +\frac{13X^2}{2m^4(\phi) } \right)\,,\label{f4hat:ap_m}
\\ \hat{F}_5(\phi,X) &\simeq&  \frac{1}{ f^2(\phi)}\left(1-\frac{X}{m^2(\phi)} -\frac{X^2}{2m^4(\phi) } \right)\,. \label{f5hat:ap_m}
\ea
Finally, upon plugging \eqref{f2hat:ap_m}-\eqref{f5hat:ap_m} back into \eqref{action_Einstein_m} and keeping terms up to $\mathcal{O}\left(X^2\right)$ the resulting Einstein-frame action reads
\be \label{eq:final_action}
S_E \simeq \int \dd^4x \sqrt{-g} \left[ \frac{\Ri}{2} - K(\phi) \frac{ (\nabla \phi)^2}{2} +  L(\phi) \frac{ (\nabla \phi)^4}{4} - U(\phi)\right]\,,
\ee
where
\be 
K(\phi) \equiv \frac{1}{f(\phi)} + \frac{U(\phi)}{m^2(\phi)} \,, \quad L(\phi) \equiv \frac{1}{m^2(\phi)} \left( \frac{1}{f(\phi)} + \frac{U(\phi)}{2 m^2(\phi)} \right) \,, \quad U(\phi) \equiv \frac{V(\phi)}{f^2(\phi)}\,.
\label{modfunc}
\ee
In the end, starting from a complicated action with involved couplings between matter and gravity, we obtained the action in terms of a single scalar degree of freedom that is minimally coupled to the Einstein-Hilbert term at the expence of a higher-order kinetic term $\propto(\nabla\phi)^4$. Note that Eq.~\eqref{eq:final_action} is almost the same as the corresponding action in the metric case~\cite{Sato:2017qau}, up to different definitions of the non-canonical kinetic functions $K(\phi)$ and $L(\phi)$. In what follows, we study the inflationary predictions of Eq.~\eqref{eq:final_action}, where we identify the scalar field $\phi$ as the inflaton of the theory. 

\subsection{Background dynamics and slow-roll}

Notice that since the gravitational sector of Eq.~\eqref{eq:final_action} is simply the Einstein-Hilbert term, the equation of motion for the connection is trivially solved by the Levi-Civita. Next, assuming that the inflaton is spatially homogeneous $\phi(x,t) = \phi(t)$, the Einstein equations turn out to be
\beq
G_{\mu\nu}=\left(K+L\dot\phi ^2\right)\nabla_{\mu}\phi \nabla_{\nu}\phi +\left(K\,\frac{\dot\phi ^2}{2} +L\,\frac{\dot\phi^4}{4}-U\right) g_{\mu\nu}\,,
\eeq
and so for the flat FLRW metric the (tt) component reads
\beq\label{eq:eom1}
3 H^2 = K\,\frac{\dot\phi^2}{2}+3L\,\frac{\dot\phi^4}{4}+U\,,
\eeq
while the scalar field equation of motion is
\beq\label{eq:eom2}
\ddot\phi \left(K+3L\dot\phi^2 \right)+3H \dot\phi \left(K+L\dot\phi^2\right)+K'\,\frac{\dot{\phi}^2}{2}+3L'\,\frac{\dot{\phi}^4}{4}+U'=0\,.
\eeq

During inflation, the $\ddot{\phi}$ term as well as the higher-order kinetic term are negligible\footnote{However, the latter may modify the dynamics during (p)reheating. We leave the study of these effects for future work. Additionally, it was shown~\cite{Tenkanen2020b} that in similar models the inflaton field goes exponentially fast to the slow-roll attractor in the very early stages of inflation and therefore the higher-order kinetic terms, that start to contribute near the end of inflation, can indeed be ignored.}. We may thus neglect them and work with the usual slow-roll approximation. 
We can make the kinetic term canonical through the field redefinition
\be \label{eq:field_trans}
\frac{\dd \chi}{\dd \phi} = \frac{\sqrt{f(\phi) + V(\phi)/m^2(\phi)}}{f(\phi)} \,.
\ee
The canonically normalized inflaton $\chi$ can in principle be obtained as a function of $\phi$ upon integration of Eq.~\eqref{eq:field_trans}. Consequently, in order to re-express the Einstein-frame action in terms of $\chi$ one should be able to invert $\chi(\phi)$ and substitute $\phi(\chi)$ into the model functions \eqref{modfunc} albeit this is not always feasible. Nevertheless, it is not necessary to work with a canonical field in order to obtain the inflationary parameters and we can circumvent this obstacle by working directly with $\phi$. To achieve this, we employ the chain rule in combination with Eq.~\eqref{eq:field_trans} in order to compute the slow-roll parameters and the number of $e$-folds as follows:
\begin{equation}
\epsilon
= 
\frac{1}{2 K} 
\left(
\frac{U'}{U}
\right)^2\,, \qquad 
\eta
= \frac{1}{U \sqrt{K}} \left( \frac{U'}{\sqrt{K}} \right)'\,, \qquad
N
= 
\int\! 
K\, \frac{U}{U'}\, \dd \phi \,,\label{eq:PSRPs}
\end{equation}
where the prime denotes differentiation with respect to $\phi$.
The number of $e$-folds at the pivot scale $k_* = 0.05\, \text{Mpc}^{-1}$ assuming instantaneous reheating~\cite{Rubio:2019ypq} is given by 
\be
N_* = 61.1 + \frac{1}{4} \log\left(\frac{U_*^2}{\rho_{\rm end}} \right)\,,
\ee
where $U_*$ is the Einstein frame potential at the pivot scale $k_*$ and $\rho_{\rm end}$ is the energy density at the end of inflation. Calculating the energy density using the method presented in \cite{Gialamas2020,Gialamas:2020snr} and assuming the limit $m^2(\phi) \rightarrow 0 $ we obtain that
\be
N_* = 60.98+\frac{1}{4} \log  U_* -\frac{1}{4} \log \left( \frac{U_{\rm end}}{U_*} \right)\,.
\ee
Now, using the fact that $U_* = \frac{3\pi ^2}{2}A_s r$  and taking into account that the term $-\frac{1}{4} \ln \left( \frac{U_{\rm end}}{U_*}\right)$ contributes insignificantly, we obtain
\be\label{eq:efolds}
N_* \simeq 56+\frac{1}{4} \ln \left(\frac{r}{0.056}\right)\,.
\ee

\vspace{0.75cm}

\section{The effect of NMDC on Higgs inflation}
\label{sec:examples}

Our main interest and motivation is the case of Higgs inflation, that is described by a non-minimal coupling function $f(\phi) = 1 + \xi \phi^2$ and a quartic potential $V(\phi) = \lambda \phi^4/4$ though we do not necessarily assume $\phi$ to be the Higgs. For these model functions, the Einstein frame potential~\eqref{modfunc} reads
\be 
U(\phi) = \frac{\lambda \phi^4}{4 \left( 1 + \xi \phi^2 \right)^2}\,.
\ee
In the following, we investigate the effect of the NMDC model function $m^2(\phi)$ on the inflationary predictions of the standard Palatini-Higgs inflation. To this end, we provide in-depth analyses for the cases of constant and quadratic couplings and briefly mention the case of a quartic coupling.

\subsection{Constant NMDC}\label{sec_Case_III}

In the simplest scenario, the coupling functional of the NMDC term will not depend on the scalar field, and thus we may write $m^2(\phi) = \kappa$. For this choice, the canonically-normalized inflaton $\chi$ can be obtained as a function of $\phi$ through the relation ~\eqref{eq:field_trans}
\be 
\frac{\dd \chi}{\dd \phi} = \frac{\sqrt{1 + \xi \phi^2 + \frac{1}{4} \frac{\lambda}{\kappa^2} \phi^4}}{1 + \xi \phi^2} \,.
\ee
The inversion of $\chi(\phi)$ can not be obtained analytically in this case and so we will work directly with the non-canonical field as we have already discussed in the previous section. The expressions for the slow-roll parameters obtained from~\eqref{eq:PSRPs} read
\be 
\epsilon = \frac{8}{\phi^2 \left( 1 + \xi \phi^2 + \frac{1}{4} \frac{\lambda}{\kappa^2} \phi^4 \right)}
\ee
and
\be 
\eta = \frac{16 \kappa^2 \left[ 4 \kappa^2 \left( 3 + \xi \phi^2 - 2 \xi^2 \phi^4 \right) + \lambda \phi^4 \left( 1 - 3 \xi\phi^2 \right) \right]}{\phi^2 \left[ \lambda \phi^4 + 4 \kappa^2 \left( 1 + \xi \phi^2 \right) \right]^2}\,.
\ee
The integral for the number of $e$-folds in this case gives
\be\label{eq:CaseIII-efolds}
N_* = \frac{\phi^2}{8} + \frac{\lambda}{32 \kappa^2 \xi^3} \left[ 2 \xi^2 \phi^4 - \xi \phi^2 + \ln \left( 1 + \xi \phi^2 \right) \right]\,.
\ee
where the first term is the usual Palatini-Higgs number of $e$-folds and the second one is attributed to $1/m^2(\phi)$. Furthermore, we have defined $N_*$ counting from $\phi = 0$. Note that the difference between this definition and the usual one where the field value at the end of inflation is obtained from $\max \left( \epsilon, \vert \eta \vert \right) = 1$ is suppressed by $1/N_*$ in the expression of $\phi(N_*)$. Next, by assuming that $\xi\phi^2\gg M_P^2$ we can further simplify the equation of $N_*$ and write the expression for the observables, now expanded around large values of $N_*$, as
\begin{align}
    A_s&\simeq \frac{N_*\left(\lambda-8\kappa\xi^2+8\sqrt{\kappa\lambda\xi^3N_*}\right)}{48\pi^2\xi^2},\\
    n_s&\simeq1-\frac{3}{2N_*}-\frac{8\kappa\xi^2-\lambda}{16N_*\sqrt{\kappa\lambda\xi^3N_*}},\\
    r&\simeq\frac{8\lambda}{N_*\left(\lambda-4\kappa\xi^2+8\sqrt{\kappa\lambda\xi^3N_*}\right)}.
\end{align}
The tensor-to-scalar ratio $r$ can also be expressed as:
\begin{equation}
    r\simeq\frac{2\lambda}{\xi^2N_*\left(\kappa+\frac{24\pi^2A_s}{N_*}\right)}
\end{equation}
and similarly the second-order correction to the spectral index, denoted hereafter as ${n_s}^{(2)}$, becomes
\begin{equation}
    {n_s}^{(2)}=-\frac{1}{2N_*}\,\left(1+\frac{48\pi^2A_s}{N_*\left(8\kappa-\frac{\lambda}{\xi^2}\right)}\right)^{-1} \,.
\end{equation}
Assuming $N_*=50$, we obtain the following equation for $r$:
\begin{equation}
    r\simeq\frac{\lambda}{\xi^2}\,\frac{4\cdot 10^{-2}}{\kappa+5\cdot10^{-9}}\,.
\end{equation}

\begin{figure}
\centering
\includegraphics[width=0.495\textwidth]{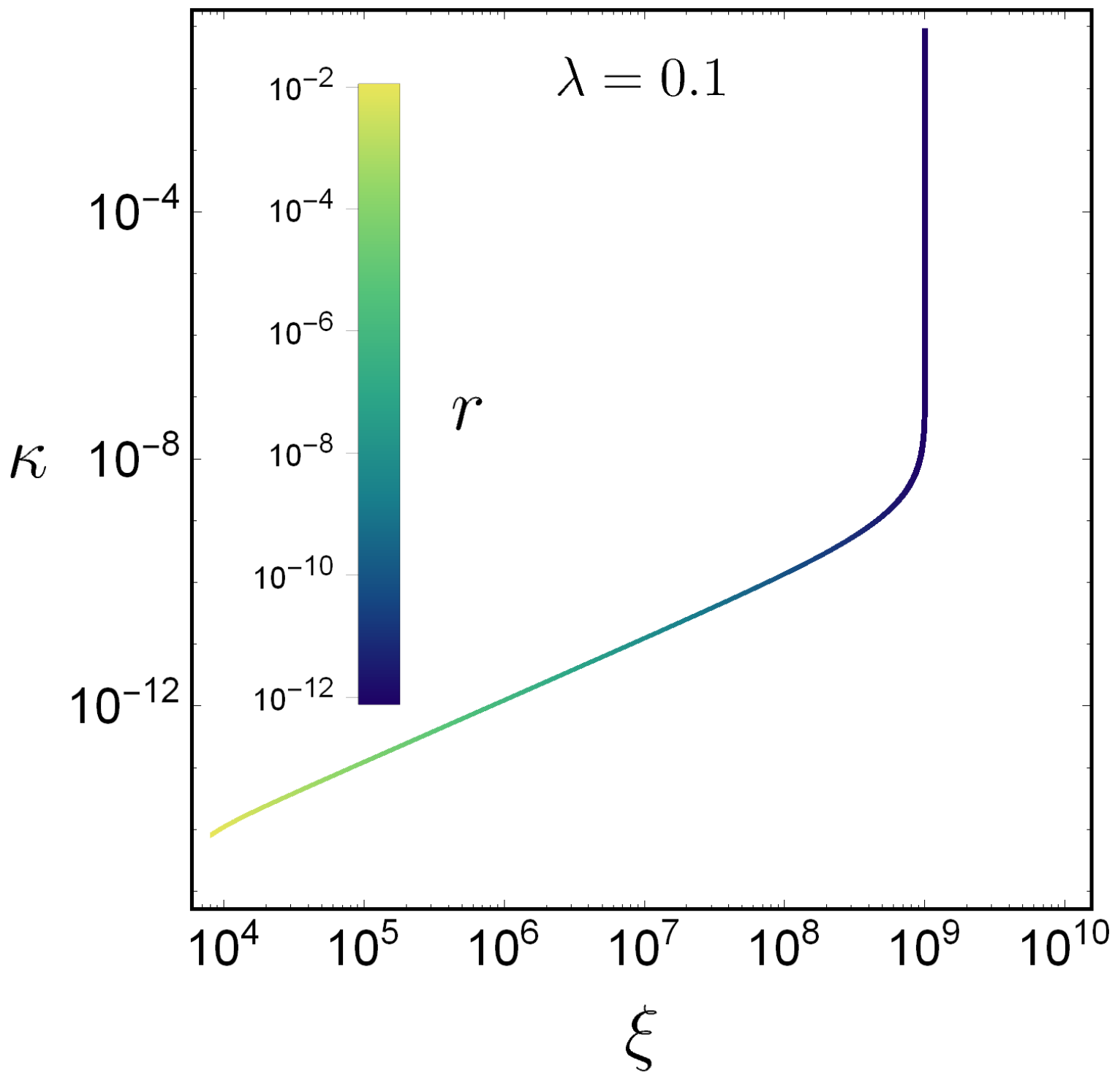}
\includegraphics[width=0.495\textwidth]{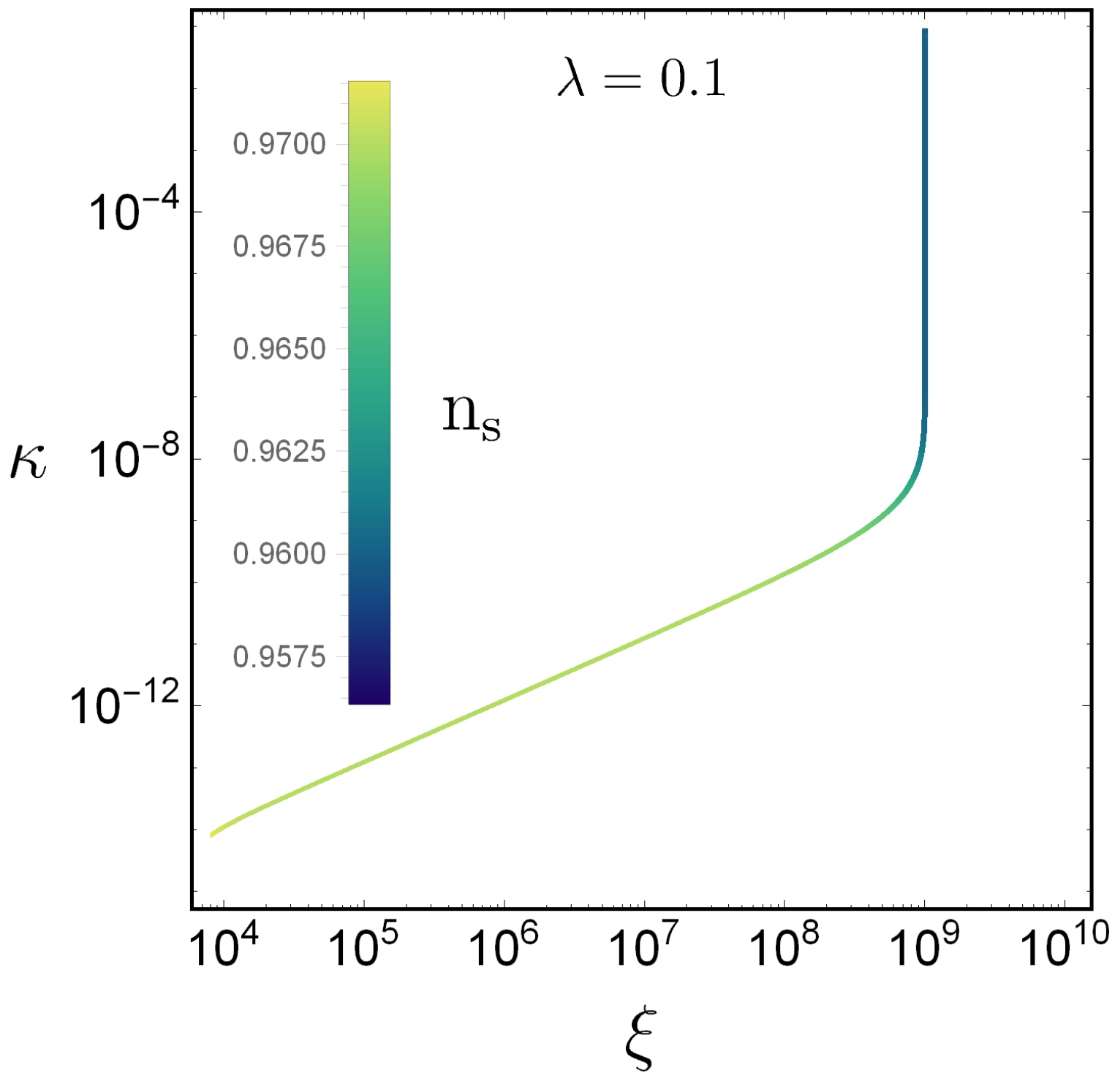}
\caption{\sf For $\lambda=0.1$ and $N_*=50$ we plot $A_s (\xi, \kappa) \simeq 2.1 \times 10^{-9}$ using the expressions presented in Appendix~\ref{App:I}. The overlaying colour-grading of the curves is associated with the corresponding values of $r$ (left) and $n_s$ (right) as they are depicted in the inlaid bars of the figures. We observe that as $\xi$ becomes smaller, we need a smaller value for $\kappa$ in order to comply with the measured value of $A_s$. At the same time, the value of $r$ grows up to $\sim 10^{-2}$, while $n_s$ grows above $\sim 0.97$. For $\xi \lesssim 10^{4}$ the validity of the $\sqrt{\xi} \phi \gg 1$ approximation fails.}
\label{fig:Case3-density-r}
\end{figure}

\textbf{Case (a)}: If $\kappa<10^{-10}$, at the marginal limit we can further simplify it as
\begin{equation}
    r\simeq 4\times10^{9}\,\frac{\lambda}{\xi^2},
\end{equation}
and in order to satisfy the $2\sigma$ bound we end up with
\begin{equation}
    \frac{\lambda}{\xi^2}\lesssim1.6\times 10^{-11}.
\end{equation}
In this case, the parameter $\kappa$ can take arbitrarily small values and the above constraint still holds, meaning that larger values of $\xi$ (with constant $\lambda$) are suppressing the tensor-to-scalar ratio. Therefore, the minimum value of $\xi$, for $\lambda=0.1$, is $\xi_\text{min}\simeq10^5$, resulting in the largest value of $r\sim10^{-3}$. Let us turn our attention to the value of $n_s^{(2)}$; it turns out that if $\kappa\gg\lambda/\xi^2$ we end up with a correction of $n_s^{(2)}\sim-10^{-6}$. A similar behaviour is obtained in the other limit where $\kappa\ll \lambda/\xi^2$. This is illustrated in Fig.~\ref{fig:Case3-density-r} where we notice that the value of $n_s$ is largely constant in that region.

\textbf{Case (b)}: In the region where $\kappa>10^{-7}$ we obtain
\begin{equation}
    r\simeq4\cdot10^{-4}\,\frac{\lambda}{\xi^2\kappa}\quad\implies\quad\frac{\lambda}{\xi^2}\lesssim1.6\times 10^{-8}\,,
\end{equation}
where once again we used the bound on $r$ in order to obtain the constraint on $\lambda/\xi^2$. In this case, we assumed marginal values of $\kappa\sim10^{-7}$. Then, if $\kappa\gg\frac{\lambda}{\xi^2}$ or even if they are of the same order of magnitude\footnote{The case of $\kappa\ll\lambda/\xi^2$ is unrealistic due to the assumption that $\sqrt{\xi}\phi\gg 1$ in the approximate expressions. In other words, the only possibility here is that $\kappa\gtrsim\lambda/\xi^2$.}, we obtain:
\begin{equation}
    {n_s}^{(2)}\simeq\left.-\frac{1}{2N_*}\,\left[1+\frac{48\pi^2A_s}{8\kappa N_*}\right]^{-1}\right|_{N_*=50}\sim-10^{-2}\,.
\end{equation}
It turns out that values of $n_s\simeq1-3/(2N_*)\sim0.97$ have important higher-order corrections which contribute significantly and can bring the spectral index in the $1\sigma$ allowed region. This is further illustrated in Fig.~\ref{fig:Case3-density-r} following a complete analysis of the exact expressions for the observable quantities.

\subsection{Field-dependent NMDC}\label{sec_Case_I}

Next let us consider the case where the prefactor of the NMDC term in the action \eqref{action1_p} depends on the inflaton. For a quadratic coupling of the form $m^2(\phi) = \phi^2/m_0^2$, the relation~\eqref{eq:field_trans} between $\phi$ and the canonically normalized $\chi$ becomes
\be 
\frac{\dd \chi}{\dd \phi} = \frac{\sqrt{1 + \left( \xi + m_0^2 \lambda/4 \right) \phi^2}}{1 + \xi \phi^2} \,.
\ee
Similarly to the case of the previous section, the solution of the above equation cannot provide us with an analytic expression for the inverted field $\phi(\chi)$ and so the slow-roll parameters are calculated once again directly in terms of $\phi$ via~\eqref{eq:PSRPs} as
\be 
\epsilon = \frac{8}{\phi^2 \left( 1 + \xi \phi^2 + \frac{1}{4} m_0^2 \lambda \phi^2   \right)} \,,
\ee 
\be 
\eta = \frac{4 \left[ 3 + \left( m_0^2 \lambda/2 + \xi \right) \phi^2 - 2 \xi \left( m_0^2 \lambda/4 + \xi \right) \phi^4 \right]}{\phi^2 \left( 1 + \xi \phi^2 + \frac{1}{4} m_0^2 \lambda \phi^2   \right)^2} \,.
\ee
The integral for the number of $e$-folds~\eqref{eq:PSRPs} can be performed exactly and it gives
\be \label{eq:CaseI_efolds}
N_*=\frac{\phi^2}{8}+\frac{m_0^2\lambda}{32\xi^2}\,\left[\xi\phi^2-\ln{(1+\xi\phi^2)}\right] \,,
\ee
where once again the first term is the usual Palatini-Higgs number of $e$-folds and the second one is attributed to $1/m^2(\phi)$. Next, we can invert Eq.~\eqref{eq:CaseI_efolds} in terms of $\phi$ in the limit $\xi \phi^2 \gg M_P^2$ to obtain
\be 
\phi^2 \simeq \frac{32 N_*}{4 + \frac{m_0^2 \lambda}{\xi}} \,.
\ee
Finally, the inflationary observables can be expressed in terms of $N_*$ as
\be \label{observ:caseI}
A_s \simeq \frac{\lambda N^2_*}{3 \pi^2 \left( m_0^2 \lambda + 4 \xi \right)} \,, \qquad n_s \simeq 1 - \frac{2}{N_*} - \frac{m_0^2 \lambda + \xi}{8 \xi^2 N_*^2}\,, \qquad r \simeq \frac{m_0^2 \lambda + 4 \xi}{2 \xi^2 N^2_*}\,.
\ee
From the measured value of $A_s$ we obtain the relation 
\begin{equation} 
	\xi \approx 4 \times 10^6 N_*^2 \lambda - \frac{m_0^2 \lambda}{4} \,.
\end{equation}
Note that in the $m_0 \rightarrow 0$ limit, the above expressions reduce to those of the standard Palatini-Higgs inflation.
One can see that if $m_0^2 \lambda$ is comparable to or larger than $4\xi$ then we can have a smaller value for the latter which translates to larger values for $r$.

\begin{figure}
\centering
\includegraphics[width=0.495\textwidth]{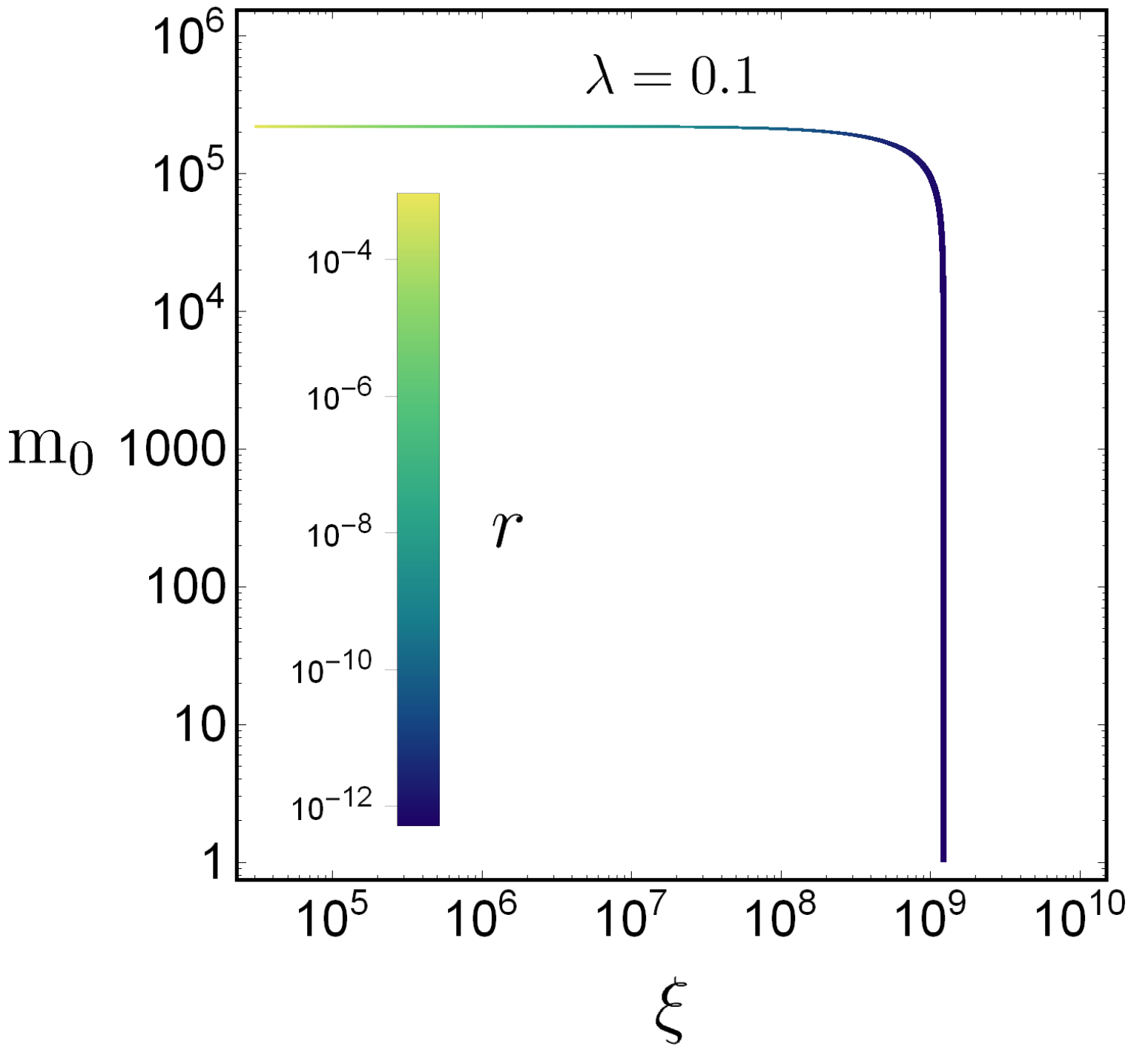}
\includegraphics[width=0.495\textwidth]{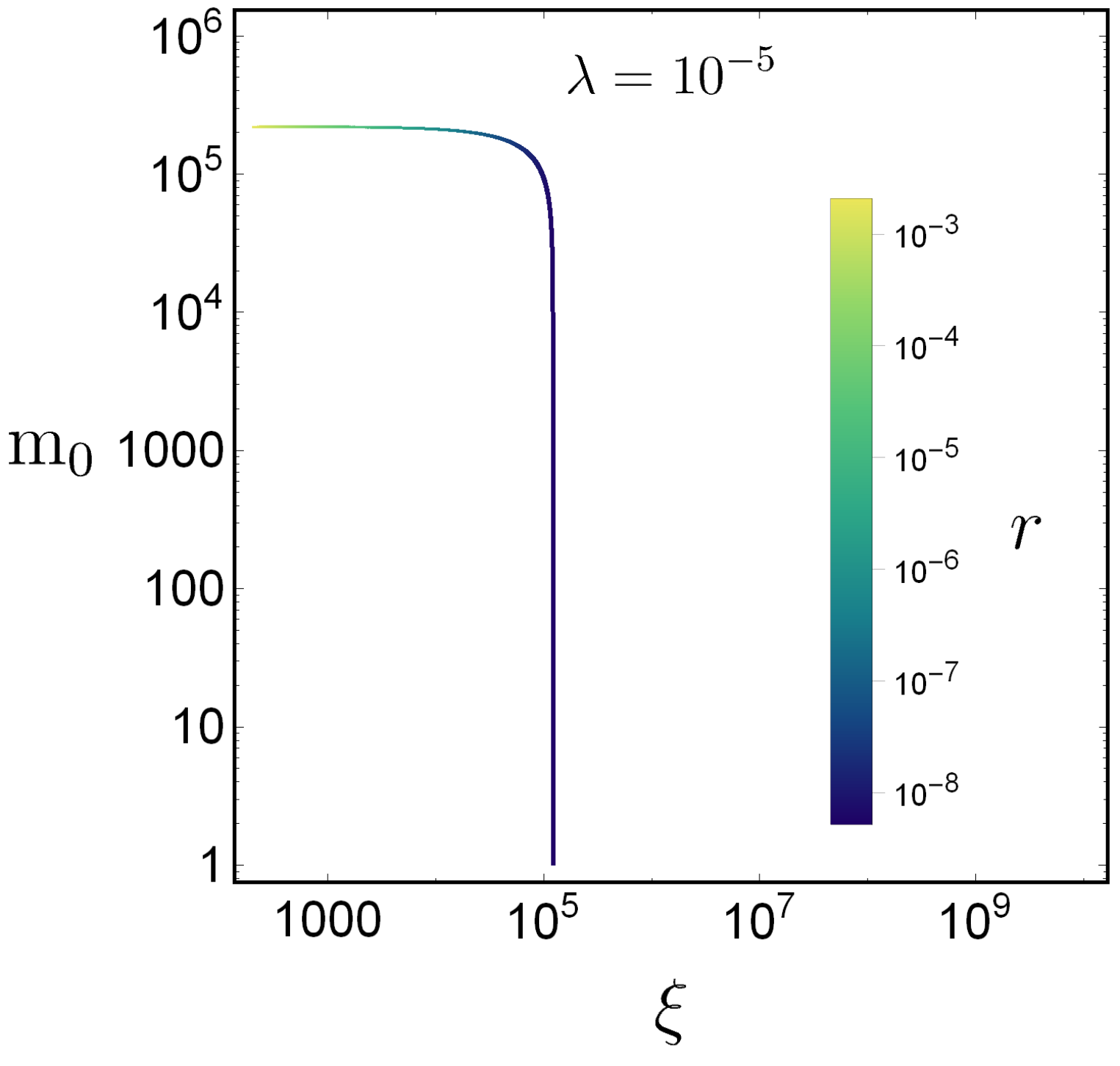}
\caption{\sf For $\lambda=0.1$ (left), $\lambda = 10^{-5}$ (right) and $N_*=55$ we plot $A_s (\xi, m_0) \simeq 2.1 \times 10^{-9}$ using the expressions presented in Appendix~\ref{App:II}. The overlaying colour-grading of the curves is associated with the corresponding values of $r$ as they are depicted in the inlaid bars of the figures. We observe that for $m_0 \simeq 2.2 \times 10^5$, the factor $m_0^2 \lambda$ in the formula~\eqref{observ:caseI} for $A_s$ dominates over $\xi$. Therefore, we can have smaller values for $\xi$, which means that $r$ can be bigger. For $\xi \lesssim 3 \times 10^{4}$ (left) and $\xi \lesssim 200$ (right) the validity of the $\sqrt{\xi} \phi \gg 1$ approximation fails.}
\label{fig:Case1-density-r}
\end{figure}

The parameters $\left\{m_0,\lambda,\xi\right\}$ have to satisfy the bound on the scalar power spectrum $A_s$, for some number of $e$-folds $N_*$. Keeping one of them constant, it is straightforward to show that the rest adjust according to the following diagram:
\begin{align}
    m_0^2=\text{const.}&\qquad\implies\qquad \xi\uparrow\,\Longleftrightarrow\,\lambda\uparrow\,,\nonumber\\
    \xi=\text{const.}&\qquad\implies\qquad m_0^2\uparrow\,\Longleftrightarrow\,\lambda\uparrow\,,\\
    \lambda=\text{const.}&\qquad\implies\qquad m_0^2\uparrow\,\Longleftrightarrow\,\xi\downarrow\,.\nonumber
\end{align}
Let us consider the tensor-to-scalar ratio r, which reads as
\begin{equation}
    r\simeq r_0\,\left(\frac{\lambda}{\xi}\,\frac{N_*^2}{12\pi^2A_s}\right),\qquad r_0\equiv \frac{2}{\xi N_*^2}\,,
\end{equation}   
where $r_0$ is the tensor-to-scalar ratio of the usual Palatini-Higgs model, presented in eq.~\eqref{eq:cmb_observables}. Assuming a modest value of number of $e$-folds $N_*=50$ we obtain the following relation:
\begin{equation}
    r\sim r_0\times\left(10^{10}\,\frac{\lambda}{\xi}\right)\,.
\end{equation}
Then, larger values of $r$ can be attained in the context of this theory, depending on the values of $\lambda$ and $\xi$. The parameter $m_0$ is eliminated in favour of $A_s$ and is therefore assumed to satisfy its observational value (together with $\xi$ and $\lambda$). Evidently, a self-coupling value of $\lambda\sim0.1$ allows for a relatively small, compared to the usual Palatini-Higgs case ($\xi \sim 10^{9}$), value of $\xi \sim 10^5$. The resulting tensor-to-scalar ratio $r \sim 10^{-4}$ lies well within the region of future experiments. This is further illustrated in Fig.~\ref{fig:Case1-density-r}, where as $\xi$ assumes smaller values and for constant $\lambda$, $r$ tends to grow.

In the case of $n_s$, the second order correction reads as:
\begin{equation}
    {n_s}^{(2)}=-\frac{r}{4}+\frac{3}{8\xi N_*^2}\simeq\left.-\frac{1}{\xi N_*^2}\left(10^{10}\,\frac{\lambda}{\xi}-\frac{3}{8}\right)\right|_{N_*=50},
\end{equation}
which is at best ${n_s}^{(2)}\sim 10^{-4}$, for $\lambda=0.1$ and $\xi=10^5$, meaning that the spectral index is largely unaffected in this case and assumes values around $n_s\sim1-2/N_*\simeq0.96$ (inside the $1\sigma$ region of observational bounds).

Closing this section, we note that a coupling function of the form $m^2(\phi)\propto\phi^4$ yields the same results as the usual Palatini-Higgs model. The main difference being that the scalar spectral index is significantly modified by higher-order contributions proportional to $m^2$, and assumes values outside the $2\sigma$ allowed region, $n_s\lesssim0.95$. In principle, one can study coupling functions of higher-order in $\phi$, for example $m^2(\phi)\propto\phi^n$, which we expect to also have large discrepancies with the observational bounds. This is, however, beyond the scope of this work.

\subsection{Numerical results}

As discussed previously, our expressions for the inflationary observables are valid assuming that $\sqrt{\xi}\phi\gg 1$. This assumption is not valid anymore for ``small" values of the non-minimal coupling, i.e.~$\xi \lesssim 10^4$ in the case where $\lambda=0.1$. For such values a numerical analysis is needed. In order to obtain the observables numerically we solve simultaneously Eqs.~\eqref{eq:eom1} and~\eqref{eq:eom2} without omitting the higher order in the velocity terms [$L(\phi)$ terms]. As we alluded to earlier, the $L(\phi)$-terms are negligible during inflation, but near its end their contribution can be increased significantly, resulting in a speed of sound that deviates from unity. In our numerical treatment the varying speed of sound (even at the end of inflation) forces us to use more accurate expressions in order to calculate the observables. For more details about the formulas for $A_s$, $n_s$ and $r$ that we have used we point the Reader to~\cite{Martin:2013uma} and~\cite{Lorenz:2008je, Lorenz:2008et, Agarwal:2008ah, Jimenez:2013xwa}.

In Table~\ref{table:1} we display sample outputs for the field-dependent and constant NMDC models. In both cases we have chosen the parameter $\xi$ in order to reach an agreement with characteristic values of the tensor-to-scalar ratio $r$ and the self-interaction coupling is fixed to $\lambda=0.1$. These characteristic values include: i) the Planck upper bound of $r=0.056$~\cite{Akrami:2018odb}, ii) the prediction of metric Higgs inflation $r \simeq 0.003$~\cite{Bezrukov2008} and iii) the expected accuracy $r \sim 10^{-4}$ of near-future experiments~\cite{Matsumura2016, Kogut_2011, Sutin:2018onu}. Constrained by~\eqref{eq:efolds}, we do not consider values for the number of $e$-folds larger than $55$. As shown in Table~\ref{table:1}, the constant NMDC model lies marginally out of the $1\sigma$ region for $n_s$~\eqref{eq:Planck_constr} for $50$ $e$-folds, unlike the field-dependent NMDC model which is within the $1\sigma$ region in a wider range of $e$-folds. Finally, note that the parameters $m_0$ and $\kappa$ are chosen in such a way that the amplitude of the power spectrum is fixed to $2.1\times 10^{-9}$ at each $N$.

\begin{table}
\begin{center}
\begin{tabular}{| c | c | c | c | c |}
\hline \hline
\multicolumn{5}{c} {Constant NMDC model }    \\      
\hline \hline  
$\xi$ & $ \kappa $ & $ N $ & $ r $ & $ n_s $ \\ \hline
 & & & & \\[-1em]
$  2.10 \times 10^{3}$ & $  2.350 \times 10^{-15}$ & $50$ & $0.0560$ & $0.9689$ \\ 
$  2.10 \times 10^{3}$ & $  1.790 \times 10^{-15}$ & $55$ & $0.0526$ & $0.9717$ \\ 
$  1.50 \times 10^{4}$ & $  1.820 \times 10^{-14}$ & $50$ & $0.0034$ & $0.9697$ \\ 
$  1.50 \times 10^{4}$ & $  1.362 \times 10^{-14}$ & $55$ & $0.0034$ & $0.9725$ \\
$ 9.00 \times 10^{4} $ & $1.135 \times 10^{-13} $ & $50$ & $0.0001$ & $ 0.9695 $ \\ 
$ 9.00 \times 10^{4} $ & $ 8.530 \times 10^{-14} $ & $55$ & $0.0001$ & $ 0.9723 $ \\
\hline \hline 
\multicolumn{5}{c} {Field-dependent NMDC model }   \\    
  \hline \hline   
 $\xi$ & $ m_0 $ & $ N $ & $ r $ & $ n_s $ \\ \hline 
  & & & & \\[-1em]
$  2.86 \times 10^{3}$ & $  2.753 \times 10^{5}$ & $50$ & $0.0560$ & $0.9637$  \\
$  2.86 \times 10^{3}$ & $  3.066 \times 10^{5}$ & $55$ & $0.0537$ & $0.9670$  \\
$  1.50 \times 10^{4}$ & $  2.070 \times 10^{5}$ & $50$ & $0.0035$ & $0.9606$ \\
$  1.50 \times 10^{4}$ & $  2.280 \times 10^{5}$ & $55$ & $0.0035$ & $0.9643$  \\
$  9.00 \times 10^{4}$ & $  1.985 \times 10^{5}$ & $50$ & $0.0001$ & $0.9592$ \\
$  9.00 \times 10^{4}$ & $  2.185 \times 10^{5}$ & $55$ & $0.0001$ & $0.9630$  \\
\hline \hline 
\end{tabular}
\caption{\sf Sample outputs of the models under consideration for the inflationary observables $N_*$, $r$ and $n_s$, for various values of the parameter $\xi$. The parameters $m_0$ and $\kappa$ are chosen in such a way that the amplitude of the power spectrum is fixed to $2.1\times 10^{-9}$ and the quartic coupling $\lambda$ is $0.1$. The chosen values of the tensor-to-scalar-ratio $r$ correspond to the largest allowed~\cite{Akrami:2018odb} value $r=0.056$, the prediction of metric Higgs inflation $r \simeq 0.003$~\cite{Bezrukov2008} and the expected accuracy $r \sim 10^{-4}$ of near-future experiments~\cite{Matsumura2016, Kogut_2011, Sutin:2018onu}. } \label{table:1}
\end{center}
\end{table}

\section{Conclusions}
\label{sec:conclusions}

The inflationary phase of the Universe is typically driven by extra degrees of freedom and in its simplest realization it is achieved by means of a single scalar field called the inflaton. The paradigmatic class of theories that naturally accommodate this scenario are the so-called scalar-tensor (ST) theories for which the two known variational principles i.e.~metric and Palatini in general yield different field equations.
One of the most attractive and economic approaches to ST theories is to place the Higgs field, thus far the only observed scalar field in Nature, in the role of the inflaton. At the same time, it has been shown that the inflationary predictions of Higgs inflation heavily depend on the choice of the variational principle employed. In the metric approach, the predicted value for the tensor-to-scalar ratio is of $\mathcal{O}(10^{-3})$ while in the case of the Palatini approach $r$ turns out to be  of $\mathcal{O}(10^{-12})$. 
In the near future, the increased accuracy of the experiments dedicated in the refinement of the measured values for the observable quantities of inflation will impose even stricter bounds with which the predictions of the various inflationary models will have to comply. More precisely, in the case of $r$ the expected accuracy in the measurement of its value will be of $\mathcal{O}(10^{-4})$. It is then clear that while the metric-variant predictions of Higgs inflation will most certainly soon be subject to falsification, the corresponding predictions of the Palatini-variant of the theory cannot be put to the test in the foreseeable future.

In this work, we have extended the action of the Higgs inflationary model \eqref{eq:SJordan} with the inclusion of a non-minimal derivative coupling term between the Einstein tensor and the first derivatives of the inflaton multiplied by an arbitrary smooth function of the field (see Eq.~\eqref{action1_p}). In order to recast the action of this theory into the Einstein frame where the inflationary observables can easily be computed we had to resort to a disformal transformation of the metric since the usual Weyl rescaling is insufficient. 
We have investigated in detail two cases for the coupling functional of the NMDC term. In the first case we considered a constant function while in the second case we assumed the coupling functional to be field dependent. 
In both cases we have showed that the predicted values for the tensor-to-scalar ratio of the Palatini-Higgs inflation in the presence of NMDC terms in the action can be rendered comparable with the corresponding values predicted by the metric variant of the standard theory and thus placed well within the range of values expected to be probed by the near-future experiments.

\acknowledgments

The research work of IDG was supported by the Hellenic Foundation for Research and Innovation (H.F.R.I.) under the ``First Call for H.F.R.I. Research Projects to support Faculty members and Researchers and the procurement of high-cost research equipment grant'' (Project Number: 824). AK was supported by the Estonian Research Council grants MOBJD381 and MOBTT5 and by the EU through the European Regional Development Fund CoE program TK133 ``The Dark Side of the Universe." The research of AL is co-financed by Greece and the European Union (European Social Fund - ESF) through the Operational Programme “Human Resources Development, Education and Lifelong Learning” in the context of the project “Strengthening Human Resources Research Potential via Doctorate Research” (MIS-5000432), implemented by the State Scholarships Foundation (IKY). TDP acknowledges the support of the grant 19-03950S of Czech Science Foundation (GA\v{C}R).

\appendix

\section{The exact slow-roll expressions for the observables}
\label{sec:appendixB}

Under the assumption that $\sqrt{\xi} \phi \gg 1$ the analytic expressions for the inflationary observables in terms of the number of $e$-folds and the parameters of the model are given below for the two choices of the NMDC functional studied in this work.

\subsection{Constant NMDC}
\label{App:I}

When $m^2(\phi)=\kappa$ the inflationary parameters turn out to be
\begin{align}
    r&=\frac{256 \kappa \lambda^2 \xi^3}{B \left[ 2 \kappa \xi^2 \left( \lambda +16 N_* \lambda \xi - B \right)+\lambda B \right]}\,,\\
    n_s - 1&=\frac{16 \kappa \lambda \xi ^3 \left\{ 64 \kappa^2 \xi^4 \left( 8 N_* \lambda \xi - B \right)-11 \lambda^2 B-2 \kappa \lambda \xi^2 \left[ \lambda \left(3+176 N_* \xi \right)+ B \left(48 N_* \xi -25 \right) \right] \right\}}{B \left[ 2 \kappa \xi^2 \left( \lambda +16 N_* \lambda \xi - B \right)+\lambda B \right]^2}\,,\\
     A_s&= \frac{B^3 \left[ 2 \kappa \xi^2 \left( \lambda +16 N_* \lambda \xi - B \right)+\lambda B \right]}{1536 \pi^2 \kappa \lambda \xi^5 \left( \lambda +B \right)^2}\,,
\end{align}
 where in order to write the expressions in a compact form we have introduced the following quantity:

\beq
B \equiv \lambda - 4\kappa \xi^2 + \sqrt{\lambda^2+16 \kappa^2 \xi^4 +8 \kappa \lambda \xi^2 \left(8 N_* \xi-1 \right)} \,.
\eeq

\subsection{Field-dependent NMDC coupling}
\label{App:II}

In the case of $m^2(\phi)=\frac{\phi^2}{m_0^2}$ the inflationary parameters are

\begin{align}
    r&=\frac{4 m_0^2 \lambda+16 \xi}{N_* \xi \left(1+8 N_* \xi \right)}\,,\\
    n_s&=\frac{4 \xi \left( 1+8 N_* \xi\right) \left[ N_*-3 +8 N_* \xi \left(N_*-2 \right) \right]-m_0^2 \lambda \left( 3+32 N_* \xi \right)}{4 N_* \xi \left(1+8 N_* \xi \right)^2}\,,\\
     A_s&= \frac{128 N_*^3 \lambda \xi^3 \left(1+8 N_* \xi \right)}{3 \pi^2 \left( m_0^2 \lambda +4 \xi \right) \left( m_0^2 \lambda +4 \xi \left( 1+8 N_* \xi \right) \right)^2}\,.
\end{align}

\bibliography{Palatini_NMDC}

\end{document}